\documentclass[review,onefignum,onetabnum]{siamart171218}

\title{A Nonlinear Transform for the Diagonalization of the Bernoulli-Laplace Diffusion Model and Orthogonal Polynomials}

\author{Chjan C. Lim \and William Pickering\thanks{Department of Mathematical Sciences, Rensselaer Polytechnic Institute, 110 8th Street, Troy, New York 12180, USA}}

\begin{document}
\nolinenumbers
\maketitle

  




\begin{abstract}
The Bernoulli-Laplace model describes a diffusion process of two
types of particles between two urns. To analyze the finite-size
dynamics of this process, and for other constructive results we
diagonalize the corresponding transition matrix and calculate
explicitly closed-form expressions for all eigenvalues and
eigenvectors of the Markov transition matrix $T_{BL}$. This is done
by a new method based on mapping the eigenproblem for $T_{BL}$ to
the associated problem for a linear partial differential operator
$L_{BL}$ acting on the vector space of homogeneous polynomials in
three indeterminates. The method is applicable to other Two Urns
models and is relatively easy to use compared to previous methods
based on orthogonal polynomials or group representations.

\end{abstract}

\section{Introduction}

The Bernoulli-Laplace (BL) model arise from diffusion theory and is related to the shuffling of cards \cite{flatto1985}. Symmetries of the
permutation group $S_{N}$ appear naturally in this model and other random
walks on groups. Previous solutions of this model have appeared in
Diaconis and Shashahani \cite{diaconis1987} and in the works of Karlin and
MacGregor \cite{KG}. Group representations are used explicitly in the first; the
derivation of a non-standard inner product or equivalently a measure for
orthogonal polynomials which are related to the eigensolutions of the BL
model appears in the second.

In this paper, we give a third way for deriving exact solutions of all the
eigenvectors of the BL model, through a nonlinear transform that
triangularizes and then diagonalizes the transition matrix $T_{BL}$. In
brief, our method associates a specific linear partial differential operator
(LPDO) $L_{BL}$ that acts on the vector space of homogeneous polynomials, $%
G, $ to the matrix $T_{BL}$. The $L_{BL}$ inherits the symmetries of the BL
model; it encodes the tri-diagonal singly-stochastic (column sums are equal
to 1 ) and anti-symmetric structure of $T_{BL}$. The components of the
(right) eigenvectors (in view of the equal column sums of $T_{BL})$ of $%
T_{BL}$ is encoded in the coefficients of the homogeneous polynomial $G.$ A
classical theory for the symmetries of such LPDOs have been formulated in
terms of the Lie algebra of symmetry operators $K$ that commutes with $%
L_{BL} $ (cf. \cite{olver}).

It turns out and we exploit in our method, that the symmetries of $L_{BL}$
appear iu the form of suitable linear and nonlinear transformations $P$ on
the independent variables $x,y,$ etc. or indeterminates of $G.$ The ease of
use of this method resides in the transparent or explicit way to find these
transformations $P.$ Our algorithm is completed by associating the
transformed LPDO, $L_{BL}^{\prime },$back to what turns out to be a
triangular matrix $T_{BL}^{\prime }$; in other words, the transformation $P$
for $L_{BL}$ encodes a similarity transformation that triangularizes $T_{BL},$
i.e., $PT_{BL}P^{-1}=T_{BL}^{\prime }$, which is then solved directly for
its eigenvalues and right eigenvectors.

\bigskip Here, we give a summary of the Urn models to which the BL model is related as
an extension. The Ehrenfest model and the Polya Urn models are two of the
early solvable models in the literature \cite{friedman}. They appear as two of the
exactly solved cases in Friedman's formulation of Urn models where
precisely one urn and balls of two colors are drawn and replaced with
additions \cite{friedman}. A dual formulation of Friedman's Urn models was introduced in a
series of recent papers \cite{pickering2015,pickering_thesis}: instead of
balls of two colors and one urn, the dual formulation uses two urns and
one-colored balls. The latter is more convenient for modelling of certain
network science models \cite{baronchelli2006}, such as
the Voter model where two balls are drawn and returned to the two urns with
prescribed probabilities that depend on the order in which they are drawn.
This is because many of the network science models are irreversible Markov
chain models \cite{bremaud} which have absorbing states. Their transition
matrices $T$, unlike $T_{BL}$ for the BL model, are not symmetric, in an
essential sense, that is, there are no non-standard inner products for $R^{N}
$ in which these matrices $T$ have a symmetric form.

Using a new method based on diagonalization of transition matrices \cite{pickering2017}, we solved exactly the eigenvectors of several well-known models, including
the Ehrenfest model, the Voter model \cite{sood,
masuda, masuda2,caccioli,rogers,baronchelli}, the Moran model for genetic
drift, and the Naming game models \cite{xie}, \cite{zhang}. Most of
these models are irreversible Markov models with absorbing states, and have
essentially non-symmetric transition matrices in the sense just mentioned.
The BL model however, is based on
two urns and balls of two colors. Thus, it is not strictly in the class
of Two Urns models to which we recently applied our method. In modifying this method so that it applies to the BL model, we will have shown that the new method is not only easy to use but also flexible in extension to new problems.

One of the main points here is the technical simplicity of uncovering the
symmetries of the above LPDOs within our method, through the explicit
appearance of the expressions $u=f(x,y)$ in the coefficients of $L.$ We give
here the LPDO $L_{V}$ for the Voter model \cite
{pickering2015,pickering_thesis} to indicate what we mean: first the
propagation equation for the transition matrix $T_{V}$ is given by 
\begin{eqnarray}
a_{j}^{(m+1)}
&=&p_{j-1}a_{j-1}^{(m)}+(1-2p_{j})a_{j}^{(m)}+p_{j+1}a_{j+1}^{(m)}
\label{mequation} \\
p_{j} &=&\frac{j(N-j)}{N(N-1)};
\end{eqnarray}
the eigen-problem for the associated LPDO $L_{V}$ $=(x-y)^{2}G_{xy}$ acting
on the homogeneous polynomial $G(x,y)=\sum_{j}c_{j}x^{j}y^{N-j}$ (which
encodes the components $c_{j}$ of the right-eigenvector of $T_{V})$ is given 
\begin{equation}
(x-y)^{2}G_{xy}=N(N-1)(\lambda -1)G,
\end{equation}
which clearly suggests the transformation $u=x-y,$ $v=y.$ Indeed this
triangularized and diagonalized the Voter model and led to its complete
solution.

\bigskip Contrast this ease of use with the fact that triangularization and
diagonalization of a given transition matrix of size $N$ has computational
complexity $O(N^{3}).$ In other words, exact integration of the Two Urns
models via diagonalization of transition matrices are nontrivial problems,
that are difficult to solve but once known, the solutions are easy to
verify. \cite{pickering2015,pickering_thesis} provides a simple method to
find such explicit diagonalization and hence all eigenvectors for a class of
transition matrices from the Two Urns models, even when their transition
matrices are essentially non-symmetric. Note that the eigenproblem and
diagonalization of symmetric matrices have a lower computational complexity.

We aim here to highlight this method's ease of use, relative to the
group representation method and the method of orthogonal polynomials.
Moreover, the BL model differs significantly from the original Two Urns subclass
of models for which our method was initially formulated. Thus, we also aim to show that, with the specific introduction of a nonlinear change of independent variables, this method can be applied to more complex models than the original class of models. Since the BL
transition matrix $T_{BL}$ is from a reversible Markov chain with a
stationary distribution \cite{bremaud}, it is non-symmetric singly-stochastic only in
a trivial sense. In other words, there exists (a difficult to find)
non-standard inner product for $R^{N},$ in which $T_{BL}$ is symmetric and
hence doubly-stochastic. Karlin and McGregor \cite{KG} using their powerful integral representation method of finding
an explicit way to symmetrize $T$ by introducing a non-standard inner
product (or orthogonal measure) into the problem, have related the right
eigenvectors of the transition matrix $T_{BL}$ of the $BL$ model to the
orthogonal polynomials called the dual Hahn polynomials. A third aim of
this paper is therefore to re-derive from the
diagonalization of $T_{BL},$ this non-standard inner product in which the
dual Hahn polynomials are an orthogonal polynomial system. Note that this
non-standard inner product, once found, yields a symmetric version of $%
T_{BL} $which is an example of a Jacobi operator that arise in the the
classical moments problem \cite{akhiezer}, and is related to orthogonal
polynomials via the Riemann-Hilbert method \cite{deift}.

The paper is organized as follows: section 3 concerns the calculation of the
right eigenvectors and eigenvalues of $T$ in closed form by method
introduced in \cite{pickering2015,pickering2017,pickering_thesis}; in view
of the fact that these right eigenvectors are not orthogonal in the usual
Euclidean inner-product, section 4 concerns the transformations needed to
calculate the orthonormal system of left eigenvectors of $T_{BL},$and also
the derivation of the non-standard inner product in which the right
eigenvectors are now orthogonal; section 5 concerns the elementary proofs,
based only on the eigenvectors and eigenvalues of $T,$ for the tight upper
and lower bounds for times to stationarity in the BL model \cite
{diaconis1987} ; section 6 concerns a numerically exact evaluation of the expression for the TV norm in these bounds using the eigenvectors and
eigenvalues of $T_{BL}$ directly, hence slightly sharper estimates for the
mixing times of the BL model.

Beyond the balanced special case of the BL problem treated in detail in this
paper, the same generating function method can be used to prove similar
tight bounds for mixing times in the other cases.

\section{Transition matrix of the BL model}

Let the transition matrix $T_{BL}$ be defined so that $(T_{BL\text{ }%
})_{ij}=Pr\{n_{w}(m+1)=i\;|\;n_{w}(m)=j\}$, so that the sum of each column
is $1$. In the general BL model for balls of two colors and two urns, $%
N_{1}, $ $N_{2},$ $N_{w},$ $N_{b}$ are fixed parameters satisfying the
constraints 
\begin{equation}
N_{w}+N_{b}=N=N_{1}+N_{2}
\end{equation}
where $N$ equals total number of balls in the model. For $i=0,...,N_{w}\leq
N_{1},$ (where by abuse of notation $i$ stands for both the row label of
transposed matrix $T_{BL}^{t}$ and the number of white balls in urn 1, $n_{w}),$
the transition probabilities are explicitly given by 
\begin{eqnarray}
p_{i} &=& \Pr \left\{ n_{w}(t+1)=i+1|\text{ }n_{w}(t)=i\right\} =\frac{\left(
N_{1}-i\right) \left( N_{w}-i\right) }{N_{1}N_{2}} \\
q_{i} &=& \Pr \left\{ n_{w}(t+1)=i-1|\text{ }n_{w}(t)=i\right\} =\frac{%
i\left( N_{b}-\left( N_{1}-i\right) \right) }{N_{1}N_{2}} \\
r_{i} &=&\Pr \left\{ n_{w}(t+1)=i|\text{ }n_{w}(t)=i\right\} =1-q_{i}-p_{i}.
\end{eqnarray}

\bigskip .

\bigskip

\section{Diagonalization - Right eigenvectors of the
general BL model}

In \cite{pickering2015,pickering_thesis}, we developed an
explicit method for exactly integrating or solving a 5-parameters subclass of
a class of Two Urns models which is parametrized by six real
parameters. Our method is based on a
relationship between certain banded stochastic
matrices $T$ (such as tridiagonal and pentadiagonal non-symmetric transition
matrices of markov chain models) and the  LPDOs acting on the vector space of homogeneous polynomials, $G(x,y)$ of finite order in two
indeterminates. The symmetries of the LPDO, $L,$ associated with a given
non-symmetric singly stochastic matrix from this solvable subclass
of the Two Urns models, are identified and used explicitly to transform from
the original indeterminates ( independent variables $x,y$ say) to suitable
new variables (such as $u=f(x,y),v=g(x,y)$). In the new variables $u,v,$ the
transformed LPDO, $L^{\prime },$ acts on the (again homogeneous of same
order as $G(x,y))$ polynomial $H(u,v).$ We have shown in \cite
{pickering2015,pickering_thesis} that the transformed eigen-problem 
\begin{equation}
L^{\prime }[H(u,v)]=\lambda (N)H(u,v)
\end{equation}
for a well-defined subclass of such Two Urns problems is equivalent
(via the inverse of the original relationship between banded matrix and
LPDO) to the eigen-problem for a triangular matrix, which can then be solved
explicitly for both right and left eigenvectors. In other words, at the end
of this brief summary, the symmetries of LPDO $L$ inherited from the
original banded stochastic matrix $T,$generate an explicit similarity
transformation, $P,$ such that 
\begin{equation}
STS^{-1}=D
\end{equation}
where $D$ is diagonal, and $S$ contains the eigenvectors of $T.$

This method can be formalized as an Algorithm as follows: Given the input of a singly stochastic transition matrix $T$ of size $N+1$,

(I) Choose a suitable homogeneous polynomial of finite degree $N$, $G$ that has the components $c_i$ of a right eigenvector of $T$ as coefficients of the monomials $x^{i}y^{j}z^{k}$; part of this choice is the number of indeterminates in $G$. For example, the Voter model of size  N (number of balls) with a transition matrix $T_{V} (N)$ which is a $N+1$ by $N+1$ real matrix, requires a homogeneous polynomial $G_{V}$ of degree $N$ in the indeterminates $x,y$ because there are two urns.

\bigskip

(II) Associate the recursion inplicit in given Markov matrix $T$ to a LPDO, $L$ which acts on the homogeneous polynomial $G$; the basic elements of this association scheme are the standard linear differential operators for increasing, decreasing and not changing the numbers of balls in each urn (which correspond in the example below to the probabilities $p,q,r$ prescibed by the transition matrix), and a set of multiplication type linear operators that correspond to shifts.

\bigskip

(III) A transformation to new independent
variables,(for instance, $u=f(x,y,z),$ $v=g(x,y,z),$ $w=h(x,y,z)$) is chosen to satisfy
two conditions: 

$(A)$ the transformed polynomial 
\begin{equation*}
H(u,v,w)=H(f(x,y,z),g(x,y,z),h(x,y,z))=G(x,y,z)
\end{equation*}
is a homogeneous polynomial of the same finite degree as $G;$

$(B)$ $u=f(x,y,z),$ $v=g(x,y,z),$ $w=h(x,y,z)$ is a transformation based on
the symmetries of $L$ (cf. \cite{olver}), that is, the
combinations $f(x,y,z),$ $g(x,y,z),$ $h(x,y,z)$ of the original variables $%
x,y,z$ appear naturally in the coefficients of the LPDO, $L.$

These conditions $(A)$ and $(B)$ are clearly not sufficient to ensure the transformed LPDO
eigenproblem

\begin{equation}
L^{\prime }[H(u,v,w)]=\lambda (N)H(u,v,w)
\end{equation}
is associated with a similar triangular matrix $T^{\prime }$ which
explicitly yields all its eigenvectors $b_{i}$. That they are sufficient has to be proved either in each problem to which we apply the Algorithm, or for a class of models as in the case of the Two Urns models. 

\bigskip

(IV) Using the transformation in step (III), derive the corresponding transformed LPDO, $L^{\prime}$ that acts on the transformed polynomial $H(u,v,w)$.

\bigskip

(V) Without explicitly calculating the transformed matrix $T^{\prime}$ which is associated with the transformed LPDO, $L^{\prime}$ in step (IV), check that the transformed eigen-problem for $L^{\prime}$ is indeed a recursion system for the transformed eigenvectors $b_{i}$ that can be solved explicitly, i.e., it is equivalent to a triangular linear system of equations. Solve for the eigenvalues and then the eigenvector components $b_{i}$, and if required transform back to the original components $c_{i}$. These are the main outputs of the Algorithm.

\bigskip

(VI) Use the eigenvectors in step (V) to diagonalize the original matrix if necessary.
This is the end of the ALgorithm.

Now we apply the Algorithm to the BL model.
Given the transition matrix of the BL model (cf. section 2), it
will be obvious that three independent variables (instead of the two before) should be used to formulate the BL problem. In step (I) of the Algorithm, we adopt the
anzatz that the $LPDO,$ $L_{BL},$ associated with the above $N$ by $N$
matrix $T_{BL}$, now acts on a homogeneous polynomial $G(x,y,z)$ in three
indeterminates, $x,y,z.$
We encode the entries $c_{k}(i)$, $%
i=0,...,N_{w}$ of the $k-th$ eigenvector of the transition matrix for the BL model 
as follows:

\begin{equation}
G^{(k)}(x,y,z)=\sum_{i=0}^{N_{w}}c_{k}(i)x^{i}y^{N_{1}-i}z^{N_{w}-i}.
\end{equation}
where $i=$ number of white
balls in urn 1 (also denoted $n_{w})$.
The choice of three independent variables to encode the components of an
eigenvector of $T_{BL}$ in the homogeneous polynomial $G(x,y,x)$ is now made
obvious by this explicit expression for $G.$

In step (II) of the Algorithm, we derive from the original eigen-problem for transition matrix $T_{BL},$ an
LPDO, $L_{BL}$, that acts on $G^{(k)}$. Towards that aim, we note, in particular, the entries
for $p_{i}$ and $q_{i}$ in $T_{BL}$ correspond respectively to the following
linear differential operators with coefficients that are monomials in $x,y,$%
and $z,$ 
\begin{eqnarray}
L_{p}= &&\frac{yzG_{yz}^{(k)}}{N_{1}N_{2}} \\
L_{q}= &&\frac{N_{b}x}{N_{1}N_{2}}G_{x}^{(k)}-\frac{xy}{N_{1}N_{2}}%
G_{xy}^{(k)},
\end{eqnarray}
where $G_{yz}^{(k)}=\frac{\partial ^{2}}{\partial y\partial z}G^{(k)}$ for
example. In addition, it is part of the association scheme that
multiplication in the LPDO (cf. \cite{olver}) by the coefficient $\frac{x}{yz}$ (resp. $%
\frac{yz}{x})$ represents down (resp. up) shifts in the index $i$ within the
discrete recursion equations of the original eigen-problem for matrix $%
T_{BL.}$ The $L_{BL}$ associated with the eigen-problem of the tridiagonal
Markov matrix $T_{BL}$is given by: 
\begin{eqnarray}
L_{BL}[G^{(k)}] &=&N_{1}N_{2}\left( \lambda _{k}-1\right) G^{(k)} \\
L_{BL}[G^{(k)}] &\equiv &\left( x-yz\right) G_{yz}^{(k)}+y\left( x-yz\right)
G_{xy}^{(k)}-N_{b}\left( x-yz\right) G_{x}^{(k)}.
\end{eqnarray}

In step(III) of the Algorithm, we note that the symmetries of $L_{BL}$ with respect to transformations of
its independent variables, is expressed in the factor $\left( x-yz\right) $
in its coefficients. This suggests the transformation to the new independent
variables 
\begin{equation}
u=x-yz,y=y,z=z.
\end{equation}
Since the transformed homogeneous polynomial is now given by 
\begin{equation}
H^{(k)}(u,y,z)=G^{(k)}(x(u,yz),y,z)=%
\sum_{i}b_{i}^{k}u^{i}y^{N_{1}-i}z^{N_{w}-i}
\end{equation}
in terms of the (new) components $b_{i}^{k}$ of the $k-th$ right
eigenvector, this transformation clearly satisfies both necessary conditions (A) and (B) in step (III) of the Algorithm. 

To prove that it is sufficient for our purpose of obtaining the eigenvalues and eigenvectors exactly and for diagonalization, we
proceed by direct calculations.

In step (IV) using the following obvious identities for the transformation of partial
derivatives 
\begin{eqnarray}
\partial _{x} &=&\partial _{u} \\
\partial _{y} &=&\partial _{y}-z\partial _{u} \\
\partial _{z} &=&\partial _{z}-y\partial _{u} \\
\partial _{xy} &=&\partial _{yu}-z\partial _{u}^{2} \\
\partial _{yz} &=&\partial _{yz}-y\partial _{yu}-z\partial _{uz}+yz\partial
_{u}^{2}-\partial _{u}
\end{eqnarray}
the transformed LPDO, $L_{BL}^{\prime }$ in $H^{(k)}$, $k=0,...,$ $N_{w}$ is 
\begin{eqnarray}
N_{1}N_{2}(\lambda _{k}-1)H^{(k)} &=&L_{BL}^{\prime }[H^{(k)}], \\
L_{BL}^{\prime }[H^{(k)}] &=&-N_{b}u\partial _{u}H^{(k)}+yu\left( \partial
_{yu}-z\partial _{u}^{2}\right) H^{(k)} \\
&&+u\left( \partial _{yz}-y\partial _{yu}-z\partial _{uz}+yz\partial
_{u}^{2}-\partial _{u}\right) H^{(k)} \\
&=&u\left( \partial _{yz}-z\partial _{uz}\right) H^{(k)}-(N_{b}+1)u\partial
_{u}H^{(k)}
\end{eqnarray}

In step (V), by reversing the derivation of the original $L_{BL}$ through the
association scheme \cite{olver}, this $L_{BL}^{\prime }$ in $H$ is shown to be
equivalent to the following triangular system for the (right) eigen-problem
of the transformed matrix $T_{BL}^{\prime }$: 
\begin{eqnarray}
&&N_{1}N_{2}(\lambda _{k}-1)b_{i}^{k} \\
&=&\left( N_{1}-i+1\right) \left( N_{w}-i+1\right) b_{i-1}^{k}-i\left(
N_{w}-i\right) b_{i}^{k}-(N_{b}+1)ib_{i}^{k}.
\end{eqnarray}
We have therefore verified the sufficiency of the transformation where $%
u=x-yz$ for triangularizing (and later diagonalizing) $T_{BL}.$ This
triangular system implies the recursion 
\begin{equation}
b_{i}^{k}=\frac{\left( N_{1}-i+1\right) \left( N_{w}-i+1\right) b_{i-1}^{k}}{%
N_{1}N_{2}(\lambda _{k}-1)+i\left( N_{w}-i\right) +(N_{b}+1)i}
\end{equation}
which can be solved directly.

For nontrivial eigensolutions for $k=0,...,N_{w},$ the denominator in $%
b_{i}^{k}$ must vanish, yielding the following exact expressions for the
eigenvalues, 
\begin{eqnarray}
\lambda _{k} &=&1-\frac{k\left( 1-k+N_{w}+N_{b}\right) }{N_{1}N_{2}} \\
&=&1-\frac{k\left( 1-k+N\right) }{N_{1}N_{2}} \\
&=&1-\frac{k\left( N-k+1\right) }{N_{1}N_{2}}
\end{eqnarray}
In the case $N_{1}=N_{w},$ 
\begin{eqnarray}
\lambda _{0} &=&1 \\
\lambda _{1} &=&1-\frac{N}{N_{1}N_{2}}.
\end{eqnarray}

The eigenvectors (in the transformed variables of $H)$ are given explicitly
by:

\begin{eqnarray}
b_{i}^{k} &=&\prod_{j=k+1}^{i}\frac{\left( N_{1}-j+1\right) \left(
N_{w}-j+1\right) }{N_{1}N_{2}(\lambda _{k}-1)+j\left( N_{w}+N_{b}+1-j\right) 
} \\
&=&\prod_{j=k+1}^{i}\frac{\left( j-N_{1}-1\right) \left( j-N_{w}-1\right) }{%
-k\left( N+1-k\right) +j\left( N+1-j\right) } \\
&=&\prod_{j=k+1}^{i}-\frac{\left( j-N_{1}-1\right) \left( j-N_{w}-1\right) }{%
\left( j-k\right) \left( j+k-N-1\right) } \\
&=&(-1)^{i-k}\frac{(k-N_{1})_{i-k}(k-N_{w})_{i-k}}{(i-k)!(2k-N)_{i-k}}.
\end{eqnarray}
Using these coefficients in the definition for $H$ gives 
\begin{equation}
H^{(k)}=\sum_{i=k}^{N^{\prime }}(-1)^{i-k}\frac{%
(k-N_{1})_{i-k}(k-N_{w})_{i-k}}{(i-k)!(2k-N)_{i-k}}%
u^{i}y^{N_{1}-i}z^{N_{w}-i}
\end{equation}

We summarize the consequences of the above steps of the Algorithm on the BL model in the following theorem:

\begin{theorem}
In the above Algorithm for the BL model, for any size $N$ of the model, the LPDO, $L_{BL}^{\prime }$, after the transformation (3.9) on the independent variables, is equivalent to a triangular linear system (3.21) which has (right) eigenvectors given by (3.31) and eigenvalues (3.25). The (right) eigenvectors of the original BL matrix $T_{BL}$ are in turn given by (3.38).
\end{theorem}

\subsection{Hypergeometric functions and dual Hahn polynomials}

\bigskip The next to final step left in this part of the paper is step (VI) in the Algorithm, to invert
the above similarity transformation to obtain explicitly the closed-form
expressions for the original components of the right-eigenvectors $c_{k}(i)$
of $T_{BL.}$ For this purpose, let $h^{(k)}(u)=H^{(k)}(u,1,1)$. Then, 
\begin{eqnarray*}
g^{(k)}(x) &=&G^{(k)}(x,1,1)=H^{(k)}(x-1,1,1)=h^{(k)}(x-1) \\
&=&(x-1)^{k}{}_{2}F_{1}(k-N_{1},k-N_{w};2k-N;1-x).
\end{eqnarray*}
Using the hypergeometric identity, 
\begin{equation}
{}_{2}F_{1}(a,b;c;1-z)\propto {}_{2}F_{1}(a,b;a+b-c+1;z)
\end{equation}
and the fact that any multiple of an eigenvector remains an eigenvector, we
take the polynomial for the right eigenvector components to be 
\begin{equation}
g^{(k)}(x)=(x-1)^{k}{}_{2}F_{1}(k-N_{1},k-N_{w};N_{2}-N_{w}+1;x)
\end{equation}
whose coefficients are the original components $c_{k}(i)$ of the $k-th$
right eigenvector corresponding to $\lambda _{k}$ prior to the
transformation above. These expressions are equivalent to the dual Hahn
polynomials \cite{KG}.

For the hypergeometric representation of the eigenvectors to be well
defined, we require $N_{2}\geq N_{w}$. There is no loss in generality with
this assumption, because we can relabel $N_{1}\leftrightarrow N_{2}$ and $%
N_{w}\leftrightarrow N_{b}$ so that the assumption holds. From the solution
for $g^{(k)}$, we expand in $x^{l},$ 
\begin{align}
g^{(k)}(x)& =\sum_{n}{\binom{k}{n}}(-1)^{k-n}x^{n}\sum_{i}\frac{%
(k-N_{1})_{i}(k-N_{w})_{i}}{(N_{2}-N_{w}+1)_{i}i!}x^{i} \\
& =\sum_{i}\sum_{n}{\binom{k}{n}}(-1)^{k-n}\frac{(k-N_{1})_{i}(k-N_{w})_{i}}{%
(N_{2}-N_{w}+1)_{i}i!}x^{i+n} \\
& =(-1)^{k}\sum_{i}\left[ \sum_{n}{\binom{k}{n}}(-1)^{n}\frac{%
(k-N_{1})_{i-n}(k-N_{w})_{i-n}}{(N_{2}-N_{w}+1)_{i-n}(i-n)!}\right] x^{i}
\end{align}
to find the explicit form for the components of the $k-th$ right
eigenvectors:. 
\begin{equation}
c_{i}^{k}=\sum_{n}{\binom{k}{n}}(-1)^{n}\frac{(k-N_{1})_{i-n}(k-N_{w})_{i-n}%
}{(N_{2}-N_{w}+1)_{i-n}(i-n)!}  \label{exact_c}
\end{equation}
Notice that the solution is the $kth$ order backwards difference of the
components of the hypergeometric coefficients of $g$.

The above treatment of
the eigen-problem by transforming via symmetries, the independent variables
of the associated LPDO, $L_{BL},$ is equivalent to a similarity
transformation of the transition matrix $T_{BL}$. Let $\mathbf{w}=\mathbf{Pv}
$ for some transformation matrix $\mathbf{P}$. Then, the eigen-problem for $%
\mathbf{w}$ is given by $\mathbf{P}\mathbf{T}_{BL}\mathbf{P}^{-1}\mathbf{w}%
=\lambda \mathbf{w}$. The above calculations is equivalent to the matrix $%
\mathbf{P}$ such that the new matrix $\mathbf{T}_{BL}^{\prime }=\mathbf{P}%
\mathbf{TP}^{-1}$ is lower triangular. The last step in this section is to
diagonalize $T_{BL.}.$ We do this by diagonalizing the matrix triangular
matrix $\mathbf{T}_{BL}^{\prime }=\mathbf{W\Lambda W}^{-1}$. Here, $\mathbf{%
\Lambda }=diag(\lambda _{0},\ldots ,\lambda _{N})$ and $\mathbf{W}$ are the
eigenvectors of $\mathbf{T}_{BL}^{\prime }$. The components of these
eigenvectors are $b_{i}$ corresponding to eigenvalue $\lambda _{k}$.
Diagonalization of $\mathbf{T}_{BL}^{\prime }$ allows us to explicitly
diagonalize the original transition matrix as 
\begin{equation}
\mathbf{T}_{BL}=\mathbf{P}^{-1}\mathbf{W\Lambda }\mathbf{W}^{-1}\mathbf{P}.
\label{diagonalization}
\end{equation}
Note that the matrix of eigenvectors is given by $\mathbf{P}^{-1}\mathbf{W}$.

\bigskip

\section{Symmetrizing transform, orthogonal measure and dual Hahn polynomials}

For transition matrix $T_{BL},$ let $Z$ be given by  (where we drop the
subscript BL herein, i.e., $T=T_{BL}$) 
\begin{equation}
Z_{ij}=\sqrt{\pi _{j}}T_{ij}\frac{1}{\sqrt{\pi _{i}}}.
\end{equation}
Recall the detailed balance of $T$ and its stationary distribution, given by 
$T_{ij}\pi _{j}=T_{ji}\pi _{i}$ \cite{bremaud} follows from the reversibility and
ergodicity of the BL model.  Note that $Z$ is the symmetric version of the
transition matrix: 
\begin{align}
Z_{ij}& =\frac{1}{\sqrt{\pi _{j}}}\pi _{j}T_{ij}\frac{1}{\sqrt{\pi _{i}}} \\
& =\frac{1}{\sqrt{\pi _{j}}}T_{ji}\sqrt{\pi _{i}} \\
& =Z_{ji}.
\end{align}
Therefore, $Z$ has an orthonormal set of left eigenvectors, $w_{k}^{T}$. Let 
$W$ be a matrix whose columns are $w_{k}$. The spectral decomposition of $Z$
by left eigenvectors is given by 
\begin{equation}
Z=W\Lambda W^{T}.
\end{equation}
By the definition of $Z_{ij}$, the transformation from $T$ to $Z$ can be
expressed as 
\begin{equation}
D^{-1}TD=Z,
\end{equation}
where $D$ is a diagonal matrix whose diagonal entries are $\sqrt{\pi _{i}}$.
So, arbitrary powers of $T$ is given by 
\begin{equation}
T^{m}=DW\Lambda ^{m}(D^{-1}W)^{T}.
\end{equation}

Defining a new transformation by $S,$%
\begin{equation}
T^{m}=S\Lambda ^{m}S^{-1}\label{e8}.
\end{equation}
\ Since $W$ has a specific normalization, we can equate $S$ with $DW$ after
applying the appropriate normalization for $S$. That is, for diagonal matrix 
$\Delta $, we take 
\begin{equation}
S\Delta =DW.
\end{equation}
We can choose any normalization for the right eigenvectors given in $S$, and 
$\Delta $ will properly renormalize them. Here, we solve for $\Delta $ and $W
$ by appealing to the orthogonality of $W$: 
\begin{equation}
W^{T}W=\Delta S^{T}D^{-2}S\Delta =I.
\end{equation}
Therefore 
\begin{equation}
S^{T}D^{-2}S=\Delta ^{-2}.  \label{e11}
\end{equation}

\bigskip \bigskip Computing the matrix multiplication on the left side
yields the diagonal entries of $\Delta $ denoted by $\Delta _{k}$ given by 
\begin{equation}
\Delta _{k}^{-2}=\sum_{i=0}^{N}\frac{1}{\pi _{i}}c_{k}(i)^{2}\label{e18}
\end{equation}
in terms of the right eigenvectors of $T_{BL}.$ Now that we have $\Delta $,
we have the representations for both the left-eigenvectors $w_{k}(i)$ and
right-eigenvectors $v_{k}(i)$ of $Z$ given by 
\begin{align}
w_{k}(i)& =\frac{\Delta _{k}}{\sqrt{\pi _{i}}}c_{k}(i) \\
v_{k}(i)& =\frac{1}{\sqrt{\pi _{i}}}w_{k}(i)=\frac{\Delta _{k}}{\pi _{i}}%
c_{k}(i)
\end{align}
in terms of the right-eigenvectors $c_{k}(i)$ of the original $T_{BL}$ that
was obtained by our method in section 2.

From Eq. \eqref{e11}, we also have an explicit formula for $S^{-1}$ given by 
\begin{equation}
S^{-1}=\Delta ^{2}S^{T}D^{-2}.
\end{equation}
So, by Eq. \eqref{e8}, we have 
\begin{equation}
T^{m}=S\Lambda ^{m}\Delta ^{2}S^{T}D^{-2}.
\end{equation}
Computing the matrix multiplication gives the following spectral
decomposition 
\begin{equation}
T_{ij}^{(m)}=\frac{1}{\pi _{j}}\sum_{k=0}^{N}\Delta _{k}^{2}\lambda
_{k}^{m}c_{k}(i)c_{k}(j).\label{e17}
\end{equation}
as the explicit representation of $Pr\{n(m)=i\;|\;n(0)=j\}$ in the BL model.

Since $T^{0}=I$, take $m=0$ in Eq. \eqref{e17} to find the stationary
distribution of the BL model 
\begin{equation}
\pi _{j}\delta _{ij}=\sum_{k=0}^{N}\Delta _{k}^{2}c_{k}(i)c_{k}(j).
\end{equation}
Note that this is the orthogonality relation for the right-eigenvectors of $%
T_{BL}$ with orthogonal measure $\Delta _{k}^{2}$ given in Eq. \eqref{e17}.
We have derived the orthogonal measure $\Delta _{k}^{2}$ in which the dual
Hahn are an orthogonal polynomial system \cite{KG}.

We summarize these results on the derivation of a non-standard inner product or orthogonal measure in which the original transition matrix of the BL model becomes a symmetric real matrix and the (right) eigenvectors are the system of orthogonal dual Hahn polynomials:

\begin{theorem}
The orthogonal measure in (4.12) symmetrizes $T_{BL}$, is related to the (left) and (right) eigenvectors of $T_{BL}$ by (4.13) and (4.14), and yields the spectral decomposition (4.17).
\end{theorem}

\bigskip

\section{Bounds of mixing times - elementary proofs}

\bigskip We will discuss first the case $N_{1}=N_{2}=N/2,$ for our method
gave the eigenvalues of the BL model to be $\lambda _{k}=1-\frac{4k\left(
N-k+1\right) }{N^{2}},$ $\lambda _{1}=1-\frac{4}{N}.$ A heuristic estimate
of the number of switches $q$ needed to mix the colors is thus, 
\begin{equation}
(1-\frac{4}{N})^{q}\simeq e^{-\frac{4q}{N}}=\frac{1}{N},
\end{equation}
and therefore $q=\frac{1}{4}N\log N$ gives an idea of how many switches or
time steps are needed until the variation distance between $\rho _{q}$ and $%
\pi $ is of order $O(1/N).$ The lower bound is obtained along the lines of
Diaconis et al, that is, by an application of the Chebyshev's inequality.
However, all estimates of the relevant mean values and variance needed to
apply the Chebyshev's inequality are constructed explicitly from the
properties of the eigenvalues and eigenvectors of the BL model. We will state the theorems below but since the proofs are similar to those in \cite{diaconis1987}, we have provided the details in an appendix.

\begin{theorem}
For $m=\frac{1}{4}N\log N+(\frac{c}{2}-\frac{\log 2}{8})N,$ for $c>0,$ there
is a universal constant $A>0$ such that $E_{\pi }\left[ \left\| \rho
_{m}(j;.)-\pi \right\| _{V}\right] \leq Ae^{-2c}.$
\end{theorem}

\bigskip

\begin{theorem}
If $m=\frac{1}{8}N\ln N-\frac{cN}{2},$ then $2||\rho _{m}-\pi ||_{V}\geq
1-e^{4c}$
\end{theorem}

\section{Exact calculations of mixing times}

Given that we can calculate $\rho_m(i,j)$ exactly by Eq. \eqref{e17},
and the stationary distribution is given by 
\begin{equation}
\pi_i=\frac{{\binom{N_w}{i}}{\binom{{N-N_w}}{{N_1-i}}}}{{\binom{N}{N_1}}},
\end{equation}
the total variational distance can be exactly computed for all time steps by 
\begin{equation}
\|\rho-\pi\|_V=\frac{1}{2}\sum_{i=0}^{N/2}\left| \pi
_{i}\sum_{k=1}^{N/2}\lambda _{k}^{m}v_{k}(i)v_{k}(0)\right|
\end{equation}
This solution is shown in Figure 1, with the upper bound given in Figure 2.

\begin{figure}[h!]
\begin{center}
\includegraphics[scale=0.40]{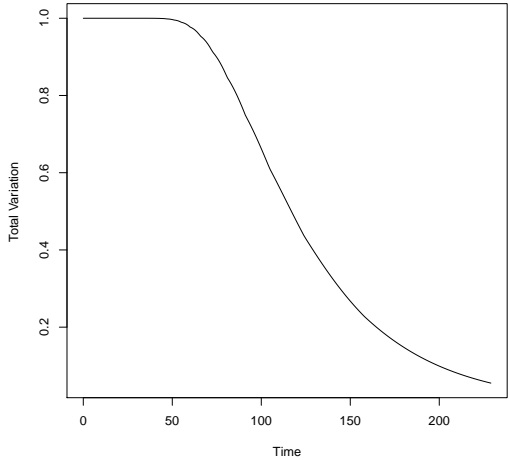}
\end{center}
\caption{Plot of the exact solution for the total variational distance for $%
N_1=N_2=N_w=N_b=100$.}
\end{figure}

\begin{figure}[h!]
\begin{center}
\includegraphics[scale=0.40]{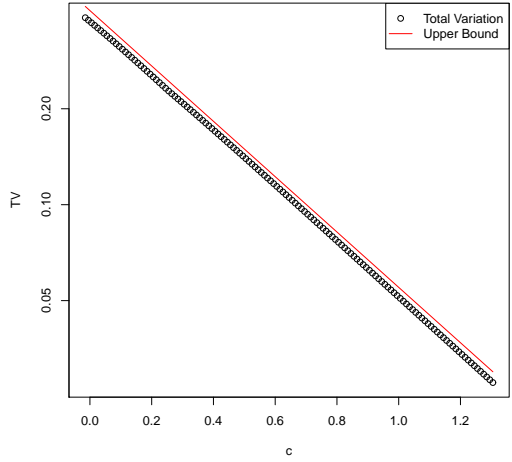}
\end{center}
\caption{Plot of the exact solution for the total variational distance, and
the upper bound given in Theorem 5.1}
\end{figure}

\newpage
\appendix
\section{Proofs of Theorems 5.1 and 5.2} 

The eigenvalues and eigenvectors of the BL problem can be used to construct
an upper bound on the variation distance between $the$ probability
distribution after $m$. As we proved in section 3, the right-eigenvectors of
the original matrix $T_{BL}$ are not orthogonal in the standard inner
product of $R^{N}$ but a different inner product weighted by $\Delta _{k}^{2}
$ can be used to derive orthogonality of a related system of 
right-eigenvectors $v_{j}$ of $T=T_{BL}.$

\subsection{Upper bound}

\bigskip\ Let $j=0$ to define 
\begin{equation*}
\rho _{m}(i;0)=\pi _{i}+\pi _{i}\sum_{k=0}^{N/2}\lambda
_{k}^{m}v_{k}(i)v_{k}(0)
\end{equation*}
where

\begin{eqnarray}
\Pr \{n_{w}(m) &=&i\text{ }|\text{ }n_{w}(0)=j\}=T_{ij}^{m}=\rho _{m}(j;i) \\
&=&\sum_{k=0}^{N/2}\pi _{i}\lambda _{k}^{m}v_{k}(i)v_{k}(j) \\
&=&\pi _{i}+\pi _{i}\sum_{k=1}^{N/2}\lambda _{k}^{m}v_{k}(i)v_{k}(j).
\end{eqnarray}
Then, $\rho _{0}(i;j)=\delta _{ij},$ and hence 
\begin{equation}
\sum_{k=1}^{N/2}v_{k}^{2}(i)=\frac{1}{\pi _{i}}-1<\frac{1}{\pi _{i}},
\end{equation}
implies that

\begin{eqnarray}
\left\| \rho _{m}(i;0)-\pi _{i}\right\| _{V} &=&\frac{1}{2}%
\sum_{i=0}^{N/2}\left| \pi _{i}\sum_{k=1}^{N/2}\lambda
_{k}^{m}v_{k}(i)v_{k}(0)\right|  \\
&\leq &\frac{1}{2}\lambda _{1}^{m}\sum_{i=0}^{N/2}\pi _{i}\left|
\sum_{k=1}^{N/2}v_{k}(i)v_{k}(0)\right|  \\
&\leq &\frac{1}{2}(N)^{-1/2}e^{-2c}\sum_{i=0}^{N/2}\pi _{i}\left(
\sum_{k=1}^{N/2}v_{k}^{2}(i)\right) ^{1/2}\left(
\sum_{k=1}^{N/2}v_{k}^{2}(0)\right) ^{1/2} \\
&\leq &\frac{1}{2}(N\text{ }\pi _{0})^{-1/2}e^{-2c}\sum_{i=0}^{N/2}(\pi
_{i})^{1/2}
\end{eqnarray}
Using 
\begin{equation}
\sum_{i=0}^{N/2}(\pi _{i})^{1/2}=O(N^{1/4})
\end{equation}
and averaging over initial data $n_{w}(0)=j,$ $j=0,...,N/2,$we have 
\begin{eqnarray}
E_{\pi }\left[ \left\| \rho _{m}(j;i)-\pi _{i}\right\| _{V}\right]  &\leq
&\sum_{j=0}^{N/2}\pi _{j}\left\| \rho _{m}(j;.)-\pi \right\| _{V} \\
&\leq &\frac{1}{2}N^{-1/2}e^{-2c}\sum_{j=0}^{N/2}(\text{ }\pi
_{j})^{1/2}\sum_{i=0}^{N/2}(\pi _{i})^{1/2} \\
&\leq &Ae^{-2c}
\end{eqnarray}
for some $A$ independent of $N.$ This proves theorem 5.1.

\bigskip
\subsection{Lower bound}

In terms of the right-eigenvectors $v_{k},$ $k=0,...,N/2,$  (with col sum =1
)

\begin{eqnarray}
\Pr \{n_{w}(m) &=&i\text{ }|\text{ }n_{w}(0)=j\}=T_{ij}^{m}=\rho _{m}(i)%
\text{ if we take }j=0 \\
&=&\sum_{k=0}^{N/2}\pi _{i}\lambda _{k}^{m}v_{k}(i)v_{k}(j) \\
&=&\pi _{i}+\pi _{i}\sum_{k=0}^{N/2}\lambda _{k}^{m}v_{k}(i)v_{k}(j)
\end{eqnarray}

and 
\begin{eqnarray}
E_{\rho _{m}}[v_{1}(i)] &=&\sum_{i=0}^{N/2}v_{1}(i)\rho
_{m}(i)=\sum_{i=0}^{N/2}\pi _{i}v_{1}(i)\sum_{k=0}^{N/2}\lambda
_{k}^{m}v_{k}(i)v_{k}(0) \\
&=&\sum_{i=0}^{N/2}\sum_{k=0}^{N/2}\lambda _{k}^{m}\pi
_{i}v_{1}(i)v_{k}(i)v_{k}(0) \\
&=&\sum_{k=0}^{N/2}v_{k}(0)\lambda _{k}^{m}\sum_{i=0}^{N/2}\pi
_{i}v_{1}(i)v_{k}(i) \\
&=&v_{1}(0)\lambda _{1}^{m}\sum_{i=0}^{N/2}\pi _{i}v_{1}(i)v_{1}(i) \\
&=&v_{1}(0)\lambda _{1}^{m}=v_{1}(0)\left( 1-\frac{4}{N}\right) ^{m} \\
E_{\pi }[v_{1}(i)] &=&0;var_{\pi }\{v_{1}(i)\}=1.
\end{eqnarray}
Next for $m=\frac{1}{8}N\log N-c\frac{N}{2}$, we get $E[v_{1}]=\frac{v_1(0)}{%
\sqrt{N}}e^{2c}$. A similar calculation gives 
\begin{align}
E_{\rho _{m}}[v_{2}(i)] =&\sum_{i=0}^{N/2}v_{2}(i)\rho
_{m}(i)=\sum_{i=0}^{N/2}\pi _{i}v_{2}(i)\sum_{k=0}^{N/2}\lambda
_{k}^{m}v_{k}(i)v_{k}(0) \\
=&\sum_{i=0}^{N/2}\pi _{i}v_{2}(i)\sum_{k=0}^{N/2}\lambda
_{k}^{m}v_{k}(i)v_{k}(0)=\sum_{i=0}^{N/2}\sum_{k=0}^{N/2}\lambda _{k}^{m}\pi
_{i}v_{2}(i)v_{k}(i)\text{ }v_{k}(0) \\
=&\sum_{k=0}^{N/2}v_{k}(0)\lambda _{k}^{m}\sum_{i=0}^{N/2}\pi
_{i}v_{2}(i)v_{k}(i)\text{ }=\text{ }v_{2}(0)\lambda
_{2}^{m}\sum_{i=0}^{N/2}\pi _{i}v_{2}(i)v_{2}(i) \\
=&v_{2}(0)\lambda _{2}^{m}\sim v_{2}(0)\left( 1-\frac{8}{N} \right) ^{m}, \\
E_{\pi }[v_{1}(i)]=&0;\text{ }var_{\pi }\{v_{1}(i)\}=1
\end{align}

Next we deduce $v_1(0)$ and $v_2(0)$ from $v_{1}^{2}=Av_2+B$, $%
v_1(i)=C(N/4-i)$, and the orthogonality of $v_i$: 
\begin{equation}
1=\sum_{i=0}^{N/2} v_1(i)^2\pi_i=\sum_{i=0}^{N/2} C^2(i-N/4)^2\pi_i\sim C^2%
\frac{N}{16}
\end{equation}
Therefore, taking $C=\frac{4}{\sqrt{N}}$, we fine $v_1(0)\sim\sqrt{N}$.
Furthermore, $Av_2(0)=v_1(0)-b$. Now, we have

\begin{align}
Var_{\rho _{m}}\{v_{1}\}=& E_{\rho _{m}}\left[Av_2+B\right] -N\lambda
_{1}^{2m} \\
=&(N-B)\lambda_2^m+B-N\lambda_1^{2m}\sim B(1-\lambda_2^m)
\end{align}
So with the same normalization as above for $v_{1},$ we deduce $%
Var\{v_{1}^{\prime }\}$ is uniformly bounded by constant $2b,$ since $%
B=b+O(\log N/N).$ Now, by Chebyschev's inequality, 
\begin{equation}
Pr_\pi\{|v_1|\leq k\}\geq 1-\frac{1}{k^2}
\end{equation}
and 
\begin{align}
Pr_{\rho_m}\{|v_1|\leq k\}&\leq Pr_{\rho_m}\{v_1\leq k\} \\
&=Pr_{\rho_m}\{E_{\rho_m}[v_1]-v_1\geq E_{\rho_m}[v_1]-k\} \\
&\leq Pr_{\rho_m}\{|E_{\rho_m}[v_1]-v_1|\geq |E_{\rho_m}[v_1]-k|\} \\
&=Pr_{\rho_m}\{(E_{\rho_m}[v_1]-v_1)^2\geq (E_{\rho_m}[v_1]-k)^2\} \\
&\leq \frac{Var_{\rho_m}(v_1)}{(E_{\rho_m}[v_1]-k)^2} \\
&\leq \frac{B}{(\sqrt{N}\lambda_1^m-k)^2}
\end{align}

Thus, if $K\subset \{0,....,N/2\}$ such that $|v_{1}|\leq k$ for $k\in K,$
we deduce 
\begin{align}
2\|\rho_m-\pi\|_V&=\sum_{i=0}^{N/2}|\rho_m(0,i)-\pi_i| \\
&\geq \sum_K |\rho_m(0,i)-\pi_i| \\
&\geq \sum_K \pi_i-\sum_K\rho_m(0,i) \\
&\geq 1-\frac{1}{k^2}-\frac{B}{(\sqrt{N}\lambda_1^m-k)^2}
\end{align}
Choose $k=d\sqrt{N}\lambda_1^m$ to obtain 
\begin{align}
2\|\rho_m-\pi\|_V &\geq 1-\left[\frac{1}{d^2}+\frac{B}{(1-d)^2}\right]%
N\lambda_1^{-2m} \\
&\geq 1-be^{4c},
\end{align}
which proves theorem 5.2.

\bibliographystyle{plain}
\bibliography{BL2018d}

\end{document}